\documentclass[a4paper,
				]{jacow}
\makeatletter%
\ifboolexpr{bool{xetex}}
{\renewcommand{\Gin@extensions}{.pdf,%
		.png,.jpg,.bmp,.pict,.tif,.psd,.mac,.sga,.tga,.gif,%
		.eps,.ps,%
}}{}
\makeatother

\ifboolexpr{bool{xetex} or bool{luatex}} % test for XeTeX/LuaTeX
{}                                      % input encoding is utf8 by default
{\usepackage[utf8]{inputenc}}           % switch to utf8

\usepackage[USenglish]{babel}
\usepackage{bm}
\ifboolexpr{bool{jacowbiblatex}}%
 {%
  \addbibresource{jacow-test.bib}
  \addbibresource{biblatex-examples.bib}
 }{}
 
\usepackage{breqn}

\begin{document}

\title{Beam-Based Diagnostics of Electric Guide Fields and Lattice Parameters for Run-1 of the Muon $\bm{g\text{-}2}$ Storage Ring at Fermilab}

\author{D. A. Tarazona\thanks{dtarazona@cornell.edu}\thanks{Currently at Cornell University.}, M. Berz, K. Makino, Michigan State University, East Lansing, MI, USA \\
		J. Mott, Boston University, Boston, MA, USA\textsuperscript{1} \\
		V. Tishchenko, Brookhaven National Laboratory, Upton, NY, USA \\
		J. D. Crnkovic, Fermi National Accelerator Laboratory, Batavia, IL, USA \\
		M. J. Syphers, Northern Illinois University, DeKalb, IL, USA\textsuperscript{1} \\
		K. S. Khaw, Shanghai Jiao Tong University, Shanghai, China \\
		J. Price, University of Liverpool, Liverpool, United Kingdom \\
		\textsuperscript{1}also at Fermi National Accelerator Laboratory, Batavia, IL, USA}
	
\maketitle

\begin{abstract}
A portion of the Muon $g\text{-}2$ Storage Ring electric system, which provides vertical beam focusing, exhibited an unexpected time dependence that produced a characteristic evolution of the stored beam during Run-1 of the Muon $g\text{-}2$ Experiment at Fermilab (E989). A method to reconstruct the Run-1 electric guide fields has been developed, which is based on a numerical model of the muon storage ring and optimization algorithms supported by COSY INFINITY. This method takes beam profile measurements from the Muon $g\text{-}2$ straw tracking detectors as input, and it produces a full reconstruction of the time-dependent fields. The fields can then be used for the reproduction of detailed beam tracking simulations and the calculation of ring lattice parameters for acceptance studies and systematic error evaluations.
\end{abstract}

\section{Introduction}

During Run-1 data collection, straw tracking detector \cite{ref:tracker_paper} measurements of the transverse muon beam revealed unexpected drifting in the beam centroid and width at early times after beam injection into the ring (i.e., $ t \lesssim 200\,\upmu\mathrm{s}$). This peculiarity was observed for all Run-1 datasets: 1a, 1b, 1c, and 1d. Furthermore, coherent betatron oscillation (CBO) frequencies of the radial centroid motion were also found to evolve during the data taking period, which slowly converged to their nominal values over the course of data taking, introducing systematic effects in the Muon $g\text{-}2$ Experiment \cite{ref:gm2_prab_run1}.

The electric guide field generated by the Electrostatic Quadrupole system (ESQ) \cite{ref:esq_jason} is utilized for vertical beam confinement, and under nominal conditions, the field (i.e. the optical lattice) becomes constant after stabilizing at $t\approx 30\,\upmu\mathrm{s}$ posterior to beam injection. In this normal scenario, CBO frequencies do not change while the stable lattice provides constant betatron tunes. Also, closed orbits are expected to be stable, and, consequently, the fixed points around which beam centroids oscillate should not drift over the data taking period. 

\begin{figure}[htbp]
  \centering
  \includegraphics[width=0.8\columnwidth]{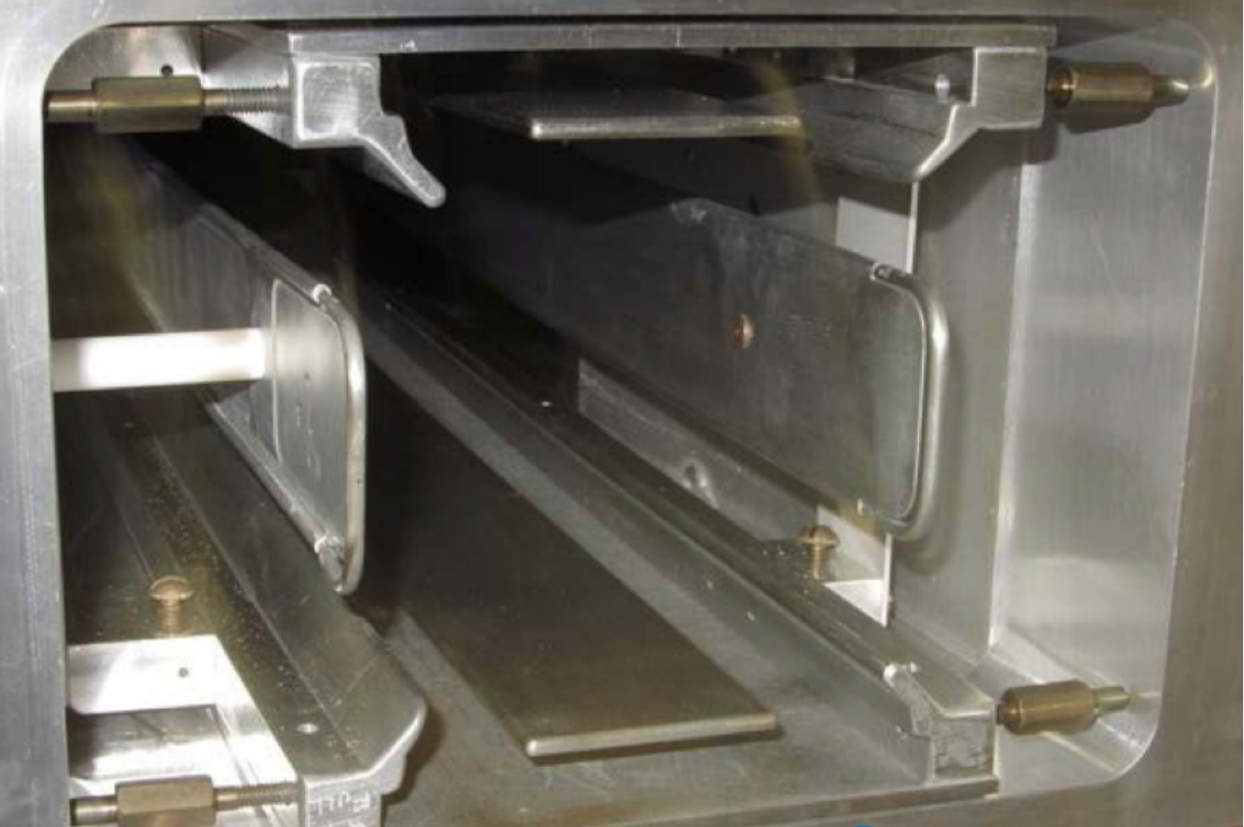}
  \caption{Photograph of one ESQ station. The top and bottom plates are held at positive voltages and the lateral plates are held at negative voltages for the vertical confinement of positive muons. The vertical magnetic field in the storage ring largely contributes to stable motion in the horizontal direction, in spite of the defocusing radial gradient from the ESQ inner and outer plates.}
  \label{fig:ESQ_pic}
  \vspace*{-\baselineskip}
\end{figure}

However, one of the eight ESQ stations (see Fig.~\ref{fig:ESQ_pic}) exhibited an unexpected behavior during Run-1. 

\begin{figure}[htbp]
  \centering
  \includegraphics[width=0.8\columnwidth]{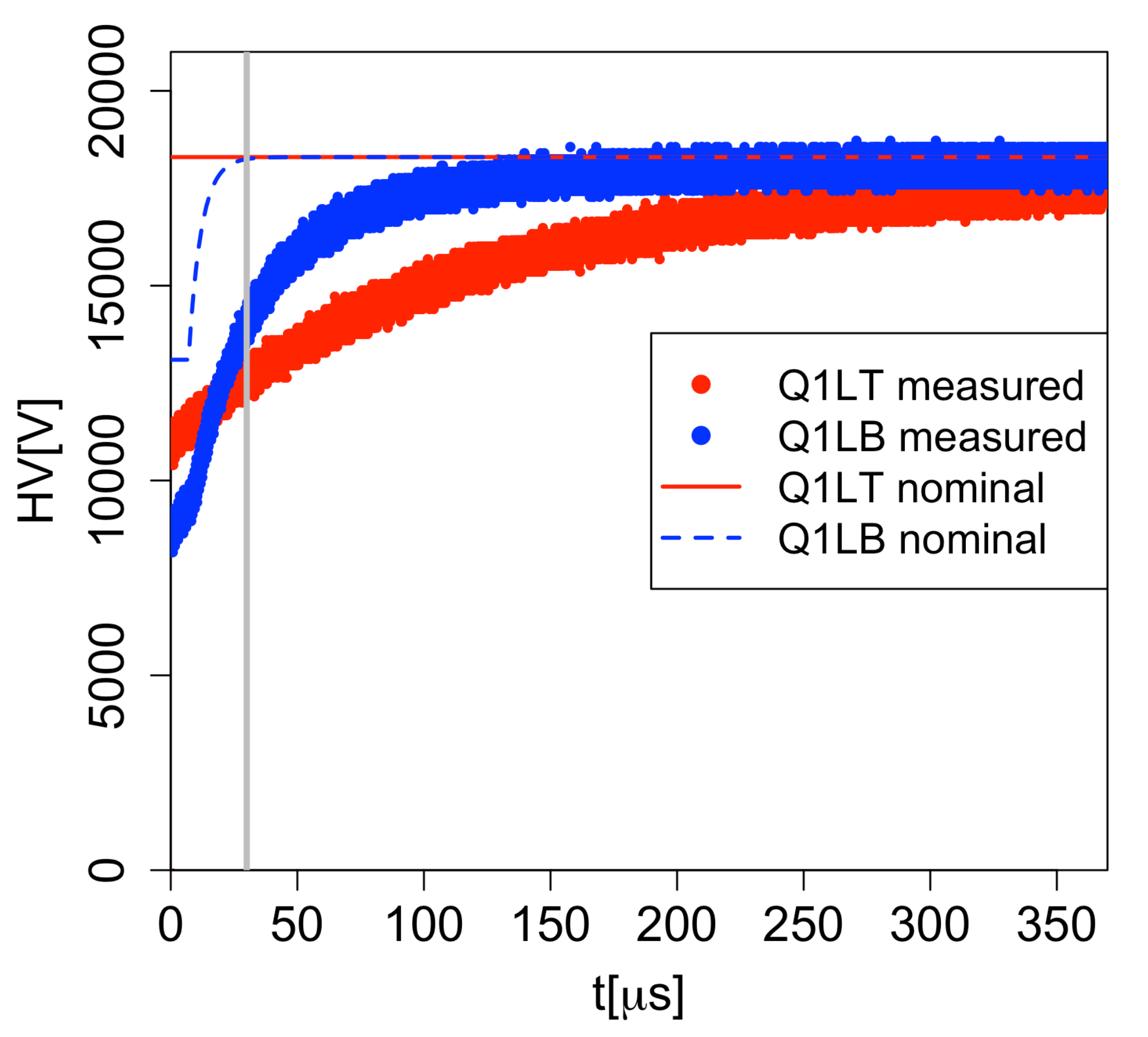}
  \caption{Sample HV traces (circle markers) from HV probe measurements in September 2018 at Q1L plates connected to faulty circuitry. Blue and red lines depict nominal HV traces.}
  \label{fig:meas_HV_18p3kV}
  \vspace*{-\baselineskip}
\end{figure}

As shown in Fig.~\ref{fig:meas_HV_18p3kV}, the high voltage (HV) applied to a top plate and a bottom plate did not follow the nominal time evolution per storing cycle. The problem was due to corona discharges on the resistors that connected these plates to the HV source; the resistors outgassed while their temperature was increasing, which would lead to discharges at low voltages. This problem was fixed prior to Run-2.

The method described in the following section was developed to ``reverse engineer'' the unmeasured HV of the misbehaving ESQ plates throughout Run-1. The ESQ station that includes these plates is commonly labeled as ``Q1L,'' which is an abbreviation that derives from its location and longitudinal dimensions within the storage ring. Based on the changing oscillation frequency of the radial centroid and vertical centroid drifts of the beam measured at the azimuthal acceptance regions of the $g\text{-}2$ straw tracking detectors, the HV of interest is reconstructed. As shown in the next section, the full ring optical functions for the Run-1 systematic-error analysis are calculated from the reconstructed HV traces and the COSY-based $g\text{-}2$ storage ring model \cite{ref:thesis_dt}.

\section{Method}

\subsection{Technique}
The electrostatic potential produced at an ESQ station can be represented as a superposition of the four contributions originated by each of its top ``T,'' bottom ``B,'' inner ``I,'' and outer ``O'' plates (see Fig.~\ref{fig:ESQ_pic}). 

In addition to the transverse potential $V_0(x,y,t)$ expected at the ESQ station of interest under nominal conditions, additional contributions, $\Delta V_T(x,y,t)$ and $\Delta V_B(x,y,t)$, from the faulty T and B plates, respectively, are included as a perturbation to the Run-1 straight-plates approximation as follows:
\begin{dmath}
\Delta V(x,y,t) = \Delta V_T(x,y,t) + \Delta V_B(x,y,t) \nonumber \\
		      = \sum_{k=0}^{\infty}\sum_{l=0}^{\infty}\left( \Delta HV_{T}(t)g_{k,l} + \Delta HV_{B}(t)b_{k,l}\right) x^{k}y^{l} \label{eq:badres_v},
\end{dmath}
where $t$ is the time passed after beam injection, $x$ is the radial distance relative to the bending radius $\rho_0=\SI{7.112}{m}$, and $y$ is the vertical distance relative to the midplane of the storage ring. In Eq.~\eqref{eq:badres_v}, $\Delta HV_{T}(t)$ and $\Delta HV_{B}(t)$ are the extra high voltages on the top and bottom plates due to the damaged resistors in Run-1 such that the total HV traces are given by 
\begin{equation}
HV_T(t)=HV_{0,T}(t)+\Delta HV_{T}(t)
\end{equation}
and
\begin{equation} 
HV_B(t)=HV_{0,B}(t)+\Delta HV_{B}(t),
\end{equation}
where $HV_{0}$ is the nominal case. The coefficients $g_{k,l}$ and $b_{k,l}$ determine the distribution of $\Delta HV_{T,B}(t)$ among the top/bottom plate multipoles. Since $b_{1,0}=0$, $b_{1,1}=0$, and $b_{k,l}=(-1)^{k+l}g_{k,l}$ due to the orientation and $180^\circ$ rotational symmetry of the top/bottom plates, Eq.~\eqref{eq:badres_v} can be rewritten as 
\begin{equation}
\Delta V(x,y,t)=\Delta V_{\mathit{dip}}^{s}(t)y+\Delta V_{\mathit{quad}}^{n}(t)(x^{2}-y^{2})+\cdots, \label{eq:vextra_expansion}
\end{equation}
where 
\begin{equation}
\Delta V{}_{\mathit{dip}}^{s}(t)=\left[\Delta HV_{T}(t)-\Delta HV{}_{B}(t)\right]b_{0,1} \label{eq:vextra_sdip_nquad1}
\end{equation}
and
\begin{equation}
\Delta V{}_{\mathit{quad}}^{n}(t)=\left[\Delta HV_{T}(t)+\Delta HV{}_{B}(t)\right]b_{2,0}. \label{eq:vextra_sdip_nquad2}
\end{equation}

The coefficients $b_{0,1}$ and $b_{2,0}$ are obtained from Ref.~\cite{ref:eremey_thesis}. Given the orthogonality of Eqs.~\eqref{eq:vextra_sdip_nquad1} and \eqref{eq:vextra_sdip_nquad2} in terms of $\Delta HV_T$ and $\Delta HV_B$, there is a unique set of top and bottom HV traces that yield $\Delta V{}_{\mathit{dip}}^{s}(t)$ and $\Delta V{}_{\mathit{quad}}^{n}(t)$. To evaluate these traces, the extra skew dipole $\Delta V{}_{\mathit{dip}}^{s}(t)$ and normal quadrupole $\Delta V{}_{\mathit{quad}}^{n}(t)$ terms must be linked to beam dynamic observables measured by the $g\text{-}2$ straw tracking detectors, as shown in the next subsection. 

\subsection{Implementation and Results}

Under the presence of the extra vertical dipole electric potential $\Delta V{}_{\mathit{dip}}^{s}(t)$ (Eqs.~\eqref{eq:vextra_expansion} and \eqref{eq:vextra_sdip_nquad1}), the vertical closed orbit becomes distorted. Therefore, by measuring the distortion of the vertical closed orbit at one azimuthal location of the storage ring (a.k.a. vertical fixed points ``$y_{0}$'') over time, $\Delta V{}_{\mathit{dip}}^{s}(t)$ can be quantified. In fact, the straw trackers have the ability to extract such vertical beam equilibrium positions around specific locations within the ring. 

To illustrate the relation between $\Delta V{}_{\mathit{dip}}^{s}(t)$ and the observable $y_{0}$ (equivalent to the non-oscillating vertical mean from tracker data), the former variable can be treated as a dipole steering error \cite{ref:mike_book}:
\begin{equation}
\left(\begin{array}{c}
{y}\\
b
\end{array}\right)_{0} = \left(I-M_{0}^{y}\right)^{-1}\left(\begin{array}{c}
0\\
\Delta\theta_{y}
\end{array}\right),
\end{equation}
\begin{equation}
\Delta\theta_{y} \approx -\frac{e\Delta{V_{\mathit{dip}}^{s}}}{E_{0}}\frac{l}{r_{\mathit{ref}}},
\label{eq:vertsteer}
\end{equation}
where $M_0^y$ is the vertical quadrant of the storage ring transfer map without the steering error, $\Delta\theta_y$ is the resulting vertical steering angle, $E_0$ the energy, $r_{\mathit{ref}}$ the reference radius of  $\Delta V{}_{\mathit{dip}}^s$, and $l$ is the length of the element that provides the steering error. It is worth mentioning that the linear vertical transfer map $M_{0}^{y}$ has to account for the gradient error.

The normal quadrupole extra term $\Delta V{}_{\mathit{quad}}^{n}(t)$ introduces a distortion to the radial defocusing gradient at Q1L. Consequently, the betatron tunes $\nu_{x,y}$ and beam transversal widths are affected due to the nonzero $\Delta V{}_{\mathit{quad}}^{n}(t)$. Trackers can indirectly measure the radial CBO frequency, $\omega_{\mathit{CBO}}$, which relates to the tunes through the cyclotron frequency, $f_{C}$, via $\omega_{\mathit{CBO}}\approx2\pi f_{C}(1-\nu_{x})$. In a similar fashion, the relation between $\Delta V{}_{\mathit{quad}}^{n}(t)$ and the observable $\omega_{\mathit{CBO}}$ can be elucidated by treating the action of $\Delta V{}_{\mathit{quad}}^{n}(t)$ as a gradient error \cite{ref:mike_book}:
\begin{dmath}
\nu_{x}=1-\frac{{{\omega}_{\mathit{CBO}}}}{2\pi f_{C}}=\frac{1}{2\pi}\left(4\cos^{-1}\left(\frac{\mathrm{Tr}(M_{0}^{x})}{2}\right)+\frac{e\Delta {V_{\mathit{quad}}^{n}}}{pv}\frac{\beta_{x}l}{r_{\mathit{ref}}^{2}}\right), \label{eq:graderr}
\end{dmath}
where $\mathrm{Tr}(M_{0}^{x})$ is the trace of the horizontal quadrant of the storage ring transfer map without the gradient error. In reality, the action of the extra skew dipole and normal quadrupole terms at Q1L during Run-1 is entangled, and magnetic field inhomogeneities already distort closed orbits. Moreover, trackers do not measure the vertical closed orbit at Q1L. Thus, the illustrative but simplistic Equations~\eqref{eq:vertsteer} and \eqref{eq:graderr} do not suffice to solve for $\Delta HV_{T,B}(t)$ via $\left( \Delta V{}_{\mathit{quad}}^{n}(t),\Delta V{}_{\mathit{dip}}^{s}(t) \right)$ with $\left(y_{0},\omega_{\mathit{CBO}}\right)$ from tracker data. For this purpose, with the high-fidelity COSY-based storage ring model \cite{ref:thesis_dt} and optimization algorithms supported by COSY INFINITY \cite{item:COSYNIMA,ref:COSYbeam}, a more representative set of bijective equations is prepared:
\begin{eqnarray}
y_{0}(t)&=F_{1}(\Delta HV_{T}-\Delta HV{}_{B},\vec{A};t),\\
\omega_{\mathit{CBO}}(t)&=F_{2}(\Delta HV_{T}+\Delta HV{}_{B},\vec{A};t).
\end{eqnarray}
With these relations fully established, the high-voltage traces are reconstructed for the top and bottom Q1L plates (a.k.a. Q1LT and Q1LB). The vector $\vec{A}$ contains all the other nominal parameters of the storage ring that are unaffected by Q1L behaviors. An iterative process to obtain the optimal top and bottom HV-trace values from tracker measurements $\left(y_{0},\omega_{\mathit{CBO}}\right)$ minimizes the objective functions ``$f_{obj}$'':
\begin{equation}
f_{obj,1}=\left(1-\frac{\nu_x^{\mathit{sim}}(\Delta HV_{T},\Delta HV{}_{B})}{\nu_x^{\text{Tracker data}}}\right)
\end{equation}
and 
\begin{equation}
f_{obj,2}=\left(1-\frac{y_{0}^{\mathit{sim}}(\Delta HV_{T},\Delta HV{}_{B})}{y_{0}^{\text{Tracker data}}}\right),
\end{equation}
where the superscript ``\textit{sim}'' stands for the values from the COSY-based model, dependent on the $\Delta HV_{T}$ and $\Delta HV{}_{B}$ values input to the model.

\begin{figure}[htbp]
	\centering
	\includegraphics[width=0.775\columnwidth]{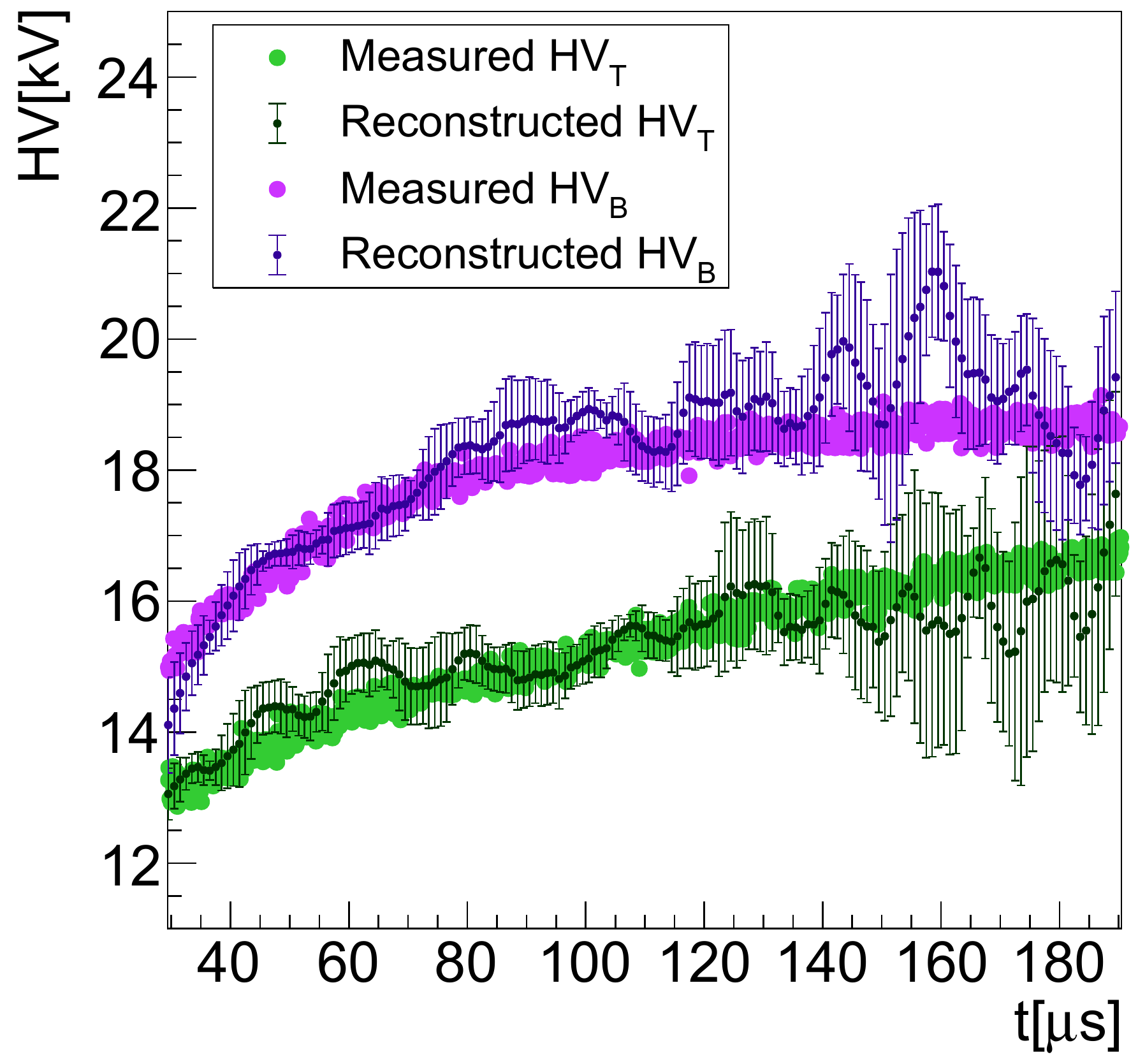} 
	\caption{Measured HV traces and reconstructed HV($t$) from a special systematic run during Run-3. The error bars are calculated based on the sensitivity of the reconstructed traces to the magnetic skew dipole term, which defines the vertical closed orbit. Fluctuations in the reconstructed values result from the tracker data statistics.}
	\label{fig:BDRun3_meas_recon}
	\vspace*{-\baselineskip}
\end{figure}

Figure~\ref{fig:BDRun3_meas_recon} shows a comparison between reconstructed HV and from direct measurements during a systematic study in year 2020, for which the damaged resistors that caused the special behavior of Q1L in Run-1 were temporarily reinserted. 

For the time window prior to the nominal measurement start time, $t<\SI{30}{\micro s}$, the HV-trace reconstruction fails to output results that resemble the functional forms as directly measured with the probe. To bypass such limitation, the functional form of the reconstructed HV traces at $t>\SI{30}{\micro s}$ is extended to fill out the gap at $t<\SI{30}{\micro s}$. The implementation of the reconstructed HV traces is validated by comparing beam tracking simulation results with tracker data, i.e., CBO frequencies and vertical centroids over time.

\begin{figure}[htbp]
	\centering
	\includegraphics[width=0.775\columnwidth]{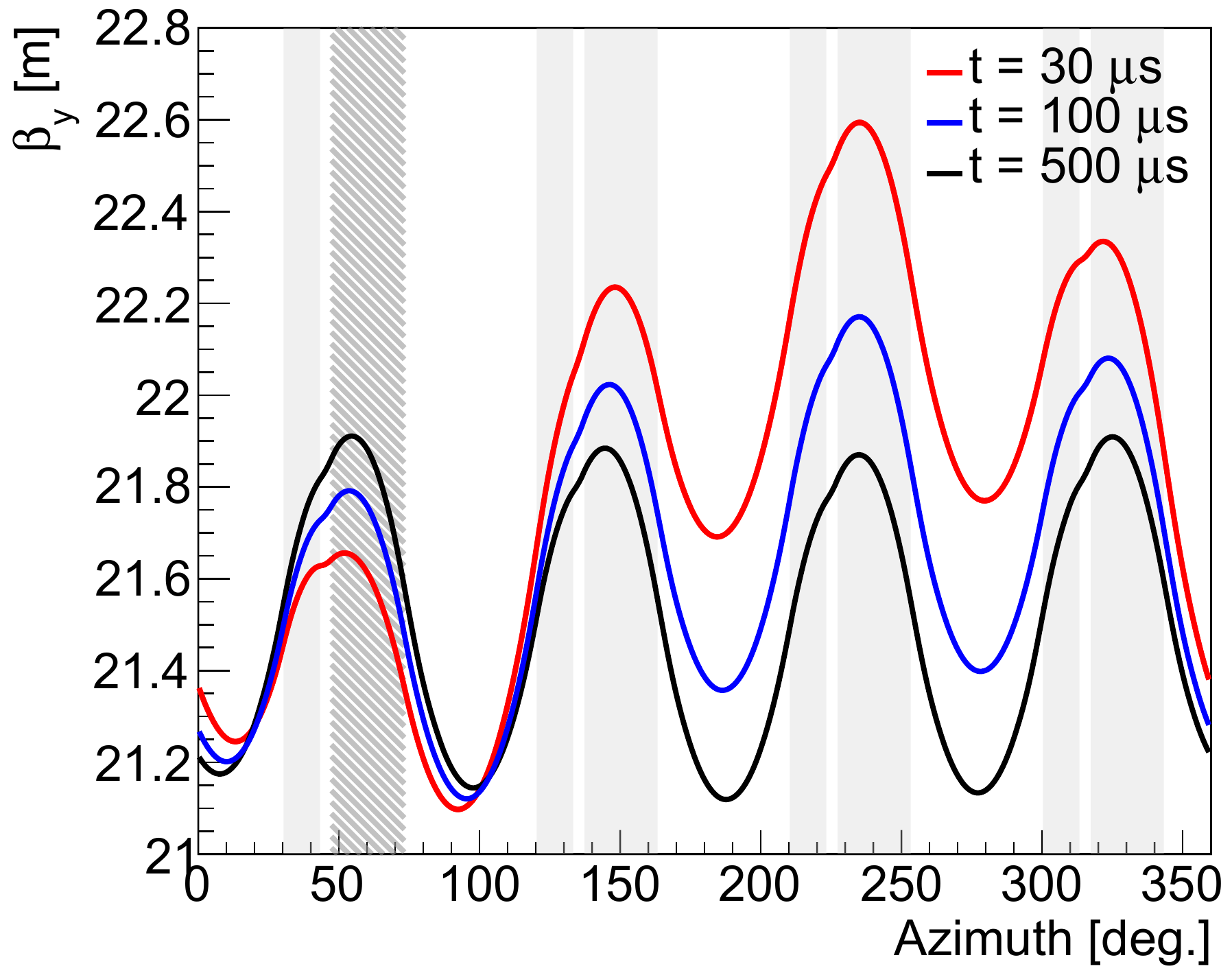} 
	\caption{Calculated vertical beta functions for Run-1 (1d) versus the ring azimuthal angle. The gray regions in the background indicate ESQ sections, where the hatched segment corresponds to Q1L.}
	\label{fig:betay_EG}
	\vspace*{-\baselineskip}
\end{figure}

With the reconstructed traces, the full electric fields of the ESQ during Run-1 are established and the time-dependent optical lattice is well defined (see Fig.~\ref{fig:betay_EG}). 

\section{CONCLUSION}

A method to reconstruct the HV traces from the observation of slowly changing beam parameters (CBO frequency and vertical mean) from beam tracker measurements in the ring was developed. Other detectors around the ring (i.e., calorimeters) provided extra inputs to constrain observations of the equilibrium vertical mean after the effect of the unstable electrode plates, this way affixing the overall vertical drifts from tracker data. Complementary simulations with independent modeling of the storage ring with the reconstructed HV traces \cite{ref:gm2_prab_run1} further validated the optical functions around the ring, and these simulations were used for the analysis of systematic effects in the Run-1 results.

\section{ACKNOWLEDGEMENTS}
Special thanks to the Muon $g\text{-}2$ Collaboration for many fruitful discussions; in particular to Bill Morse for his exceptional ideas that contributed to the method development. This work was supported by the US Department of Energy under Contract No. DE- FG02-08ER41546,  Contract No. DE-SC0018636 and prepared using the Fermi National Accelerator Laboratory (Fermilab) resources, a US Department of Energy, Office of Science, HEP User Facility. Fermilab is managed by Fermi Research Alliance, LLC (FRA), acting under Contract No. DE-AC02-07CH11359.

\bibliography{reference}

%
% only for "biblatex"
%
\ifboolexpr{bool{jacowbiblatex}}%
	{\printbibliography}%
	{%
	% "biblatex" is not used, go the "manual" way
	
	%\begin{thebibliography}{99}   % Use for  10-99  references
	
} % end \ifboolexpr
%
% for use as JACoW template the inclusion of the ANNEX parts have been commented out
% to generate the complete documentation please remove the "%" of the next two commands
% 
%%%\newpage

%%%\include{annexes-A4}

\end{document}